\documentclass[a4paper, amsfonts, amssymb, amsmath, reprint, showkeys,  twoside,twocolumn,superscript address]{revtex4-1}
\usepackage[english]{babel}
\usepackage[utf8]{inputenc}
\usepackage[colorinlistoftodos, color=green!40, prependcaption]{todonotes}
\usepackage{comment}
\usepackage{bbm}
\usepackage{amsthm}
\usepackage{mathtools}
\usepackage{physics}
\usepackage{xcolor}
\usepackage{graphicx}
\usepackage[normalem]{ulem}
\usepackage[left=23mm,right=13mm,top=35mm,columnsep=15pt]{geometry} 
\usepackage{adjustbox}
\usepackage{placeins}
\usepackage[T1]{fontenc}
\usepackage{lipsum}
\usepackage{csquotes}
\usepackage{xspace}
\newcommand{\vct}[1]{\mathbf{#1}}
\usepackage[pdftex, pdftitle={Article}, pdfauthor={Author}]{hyperref} 
\DeclareSymbolFont{bbgreek}{U}{bbold}{m}{n}
\DeclareMathSymbol{\bbmu}{\mathbb}{bbgreek}{'26}
\DeclareMathSymbol{\bbeps}{\mathbb}{bbgreek}{'17}
\newcommand{\EPF}{MF\xspace}
\newcommand{\EPFname}{\emph{motive forces\xspace}}
\newcommand{\SI}{(see SI)}

\begin{document}
\title{Near field propulsion forces from nonreciprocal media}

\author{David Gelbwaser-Klimovsky}
    \email[Correspondence email address: ]{dgelbi@mit.edu}
    \affiliation{Physics of Living Systems, Department of Physics,Massachusetts Institute of Technology, Cambridge, MA 02139, USA}
\author{Noah Graham}
\affiliation{Department of Physics, Middlebury College, Middlebury, VT 05753  USA}
\author{Mehran Kardar}
      \affiliation{Department of Physics, Massachusetts Institute of Technology, Cambridge, MA 02139, USA}
\author{Matthias Kr\"uger}
     \affiliation{
     Institute for Theoretical Physics, Georg-August-Universität, 37077 Göttingen, Germany
     }

\begin{abstract}

Arguments based on symmetry and thermodynamics may suggest the existence of a ratchet-like lateral Casimir force between two plates at different temperatures and with broken inversion symmetry. We find that this is not sufficient, and at least one plate must be made of nonreciprocal material. This setup operates as a heat engine by transforming heat radiation into mechanical force. Although the ratio of the lateral force to heat transfer in the near field regime diverges inversely with the  plates separation, $d$, an Onsager symmetry, which we extend to nonreciprocal plates, limits the engine efficiency to the Carnot value $\eta_c$. The optimal velocity of operation in the far field is of the order of $c\eta_c$, where $c$ is the speed of light. In the near field regime, this velocity can be reduced to the order of $\bar\omega d \eta_c$, where $\bar\omega$ is a typical material frequency.
\end{abstract}

\maketitle

\section{Introduction}

Fluctuations of electromagnetic fields lead to a variety of phenomena, from Planck's law of thermal radiation~\cite{Planck}, to the  fluctuation-induced normal forces predicted by Casimir in 1948~\cite{casimir1948attraction} and later generalized by Lifshitz~\cite{lifshitz1956zh}. In 1971, Polder and van Hove discussed near field effects in radiative heat transfer  between closely spaced objects~\cite{polder1971theory}, demonstrating a dramatic increase from the far field.  
These effects where observed experimentally for normal forces at the end of the last century~\cite{Lamoreaux97,Mohideen98}, and roughly ten years later for radiative heat transfer~\cite{shen2009surface,Rousseau09,biehs2020near}. A variety of related effects arise for Casimir forces in systems out of equilibrium~\cite{antezza2008casimir, Bimonte09,Messina,Kruger11, Messina11b, kruger2012trace, narayanaswamy2014a,bimonte2017nonequilibrium}, such as torques~\cite{Reid17,guo2020single} and propulsion forces~\cite{Muller15}, in both near and far field regimes.  

Lateral Casimir forces, which can propel plates with respect to each other, were also predicted 
between corrugated plates in equilibrium~\cite{Golestanian97Lateral,Emig01Lateral},
and observed experimentally~\cite{Mohideen02Lateral,Mohideen14Lateral}.
However, in these and several other 
equilibrium and non-equilibrium settings~\cite{ashourvan2007noncontact,
chiu2010lateral,Muller15,PhysRevB.97.201108},
either the shape or rotation of an object establishes the direction of the propulsive force. 
In contrast, we consider here a \textit{translationally invariant object},
and demonstrate that a lateral (potentially propulsive) force can arise by taking advantage of radiative heat transfer involving \textit{non-reciprocal} materials.
Nonreciprocal media have indeed been shown to give rise to a variety of other interesting phenomena~\cite{zhu2016persistent,herz2019green,latella2017giant,ben2016photon}.

In this Letter, we  demonstrate that  non-equilibrium propulsive forces can be used to  build a heat engine without any contact between its parts and whose optimal operation strongly depends on the type of radiation driving its heat transfer. In the far-field limit, the on-shell photon energy-momentum relation  bounds the ratio of propulsive force to heat transfer, requiring operational velocities of  the order of the speed of light, limiting the usefulness of such an engine.  In the near field, the on-shell relation does not hold, allowing the engine to operate at efficiencies close to the Carnot bound at much lower velocities.

\begin{figure}
    \centering
    \includegraphics[width=0.4\textwidth]{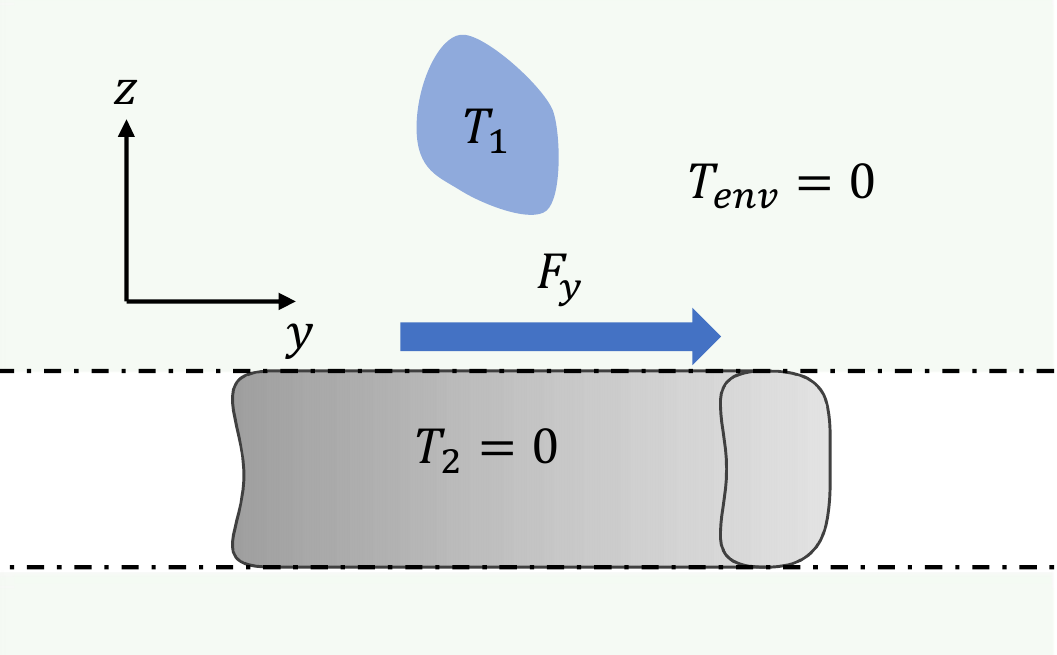}
    \caption{Two objects placed in vacuum. Object 1 is held at temperature $T_1$, while object 2 and  the environment are held at zero temperature, for simplicity. Object 2 is translationally invariant in direction $y$, and we consider the force acting on it in that direction.}
    \label{fig:1}
\end{figure}

Consider the setup depicted in Fig.~\ref{fig:1}, with two objects  held at different temperatures,  characterized by reciprocal or non-reciprocal dielectric properties encoded in their scattering operators $\mathbb{T}_i$, where $i=1,2$. We are interested in propulsive or \EPFname~(\EPF) acting on object 2 in the $y$-direction, along which it is   translationally invariant. Such forces that may be used to drive an engine are ruled out for systems at thermal equilibrium, because we can then define a Casimir free energy, which does not change under a displacement of object 2 in the $y$ direction.

  For  simplicity, let the temperatures of object 2 and of the environment be zero (this restriction will be relaxed below).  The $y$-component  of the force acting on object $2$, $F_y$, can be derived using the techniques of Ref.~\cite{kruger2012trace}, adapted to non-reciprocal media \SI, and since $\mathbb{T}_2$ is not a function of $y$, we find 
    \begin{align}
F_{y}
&=\!\frac{-2\hbar}{\pi}\! \int_0^\infty \!\!\!\!\! d\omega \frac{1}{e^{\frac{\hbar\omega}{k_BT_1}}-1} \mbox{Tr} \left\{i\partial_y{\mathbb{R}}_2\mathbb{W}\mathbb{R}_1\mathbb{W}^\dagger\right\}\!.\label{eq:tr}
\end{align}
Here we have introduced the radiation operators 
$\mathbb{R}_1=\mathbb{G}_0\left[\frac{\mathbb{T}_1-\mathbb{T}_1^\dagger}{2i} - \mathbb{T}_1 \Im[\mathbb{G}_0]\mathbb{T}_1^\dagger\right]\mathbb{G}^*_0$, $\mathbb{R}_2=\mathbb{G}_0^*\left[\frac{\mathbb{T}_2-\mathbb{T}_2^\dagger}{2i} - \mathbb{T}_2^\dagger \Im[\mathbb{G}_0]\mathbb{T}_2\right]\mathbb{G}_0$
and  the multiple scattering operator $
\mathbb{W}= \mathbb{G}_0^{-1}(1-\mathbb{G}_0\mathbb{T}_1\mathbb{G}_0\mathbb{T}_{2})^{-1}$, where $\mathbb{G}_0$ is the free Green's function. The trace in Eq.~\eqref{eq:tr} is understood to be taken over spatial coordinates as well as the indices of the $3\times3$ matrix~\cite{kruger2012trace}.
Due to the translational invariance of object 2, $\mathbb{R}_2(y,y')={\mathbb{R}}_2(y-y')=\int \frac{dk_y}{2\pi} \tilde{\mathbb{R}}_2(k_y)e^{ik_y (y-y')}=\int \frac{dk_y}{2\pi} \hat{\mathbb{R}}_2$ can be decomposed  in Fourier modes,
and the force is written 
\begin{align}
F_y =\frac{2\hbar}{\pi} \int_0^\infty d\omega \frac{1}{e^{\frac{\hbar\omega}{k_BT_1}}-1} \int \frac{dk_y}{2\pi} k_y S(k_y).\label{eq:sforcetr3}
\end{align}
Importantly,  $S(k_y)=\mbox{Tr} \{\hat{\mathbb{R}}_2\mathbb{W}\mathbb{R}_1\mathbb{W}^\dagger\}$ is the \emph{heat flux} density per wavevector $k_y$, and indeed the energy $H$ absorbed by object 2 per unit time is given by 
\begin{align}
H&= \frac{2\hbar}{\pi} \int_0^\infty d\omega \frac{\omega}{e^{\frac{\hbar\omega}{k_BT_1}}-1} \int \frac{dk_y}{2\pi} S(k_y).\label{eq:heat}
\end{align}
The  objects are made of materials described by  dielectric permittivity and magnetic permeability tensors $\bbeps$ and $\bbmu$, expressed in the potential
$\mathbb{V}=\frac{\omega^2}{c^2}(\bbeps-\mathbb{I})+\boldsymbol{\nabla}\times(\mathbb{I}-\frac{1}{\bbmu})\boldsymbol{\nabla}\times$~\cite{kruger2012trace}. For passive materials, $(\mathbb{V}-\mathbb{V}^\dagger)/i\geq0$, which are local, $\mathbb{V}_2\sim\delta(y-y')$, 
one can prove that $\hat{\mathbb{R}}_2$ is positive semidefinite for any $k_y$ \SI, and 
\begin{align}
S(k_y)\geq 0.
\label{eq:pos}
\end{align}
Equation~\eqref{eq:pos} implies that the heat flux is non-negative for any $k_y$ (note that beyond passivity we have made no assumptions regarding the shape or properties of object 1). It also tells us that the force in Eq.~\eqref{eq:sforcetr3} is due to absorbed photons, which contribute the $y$ component of their momentum to the force. This property of \EPF,  displayed in Eq.~\eqref{eq:sforcetr3}, tightly links \EPF and radiative energy  transfer. This link, which  is not present for  other Casimir forces, has physical consequences: It leads, via Eq.~\eqref{eq:heat}, to a {\it bound} on the force from heat transfer, and also places restrictions on \EPF derived from thermodynamics, as discussed below. 

In practice, the integral in Eq.~\eqref{eq:sforcetr3} is often cut off by a maximal value $k^{max}_y$ (see below). With it, the spectral densities for $F_y$ and $H$, $F_y \equiv \int_0^\infty d\omega f(\omega)$ and $H \equiv \int_0^\infty d\omega h(\omega)$,  obey
\begin{align}\label{eq:bound}
|f(\omega)|c\leq h(\omega) \frac{c}{\omega} k_y^{max}. 
\end{align}
For on-shell modes, which are dominant in the far field, $k_y^{max}=\omega/c$, so    
\begin{align}\label{eq:ff}
|f(\omega)|c\leq h(\omega),
\end{align}
illustrating that \EPF are bounded by the photon energy-momentum relation in this limit. 

 \begin{figure}
    \centering
    \includegraphics[width=0.4\textwidth]{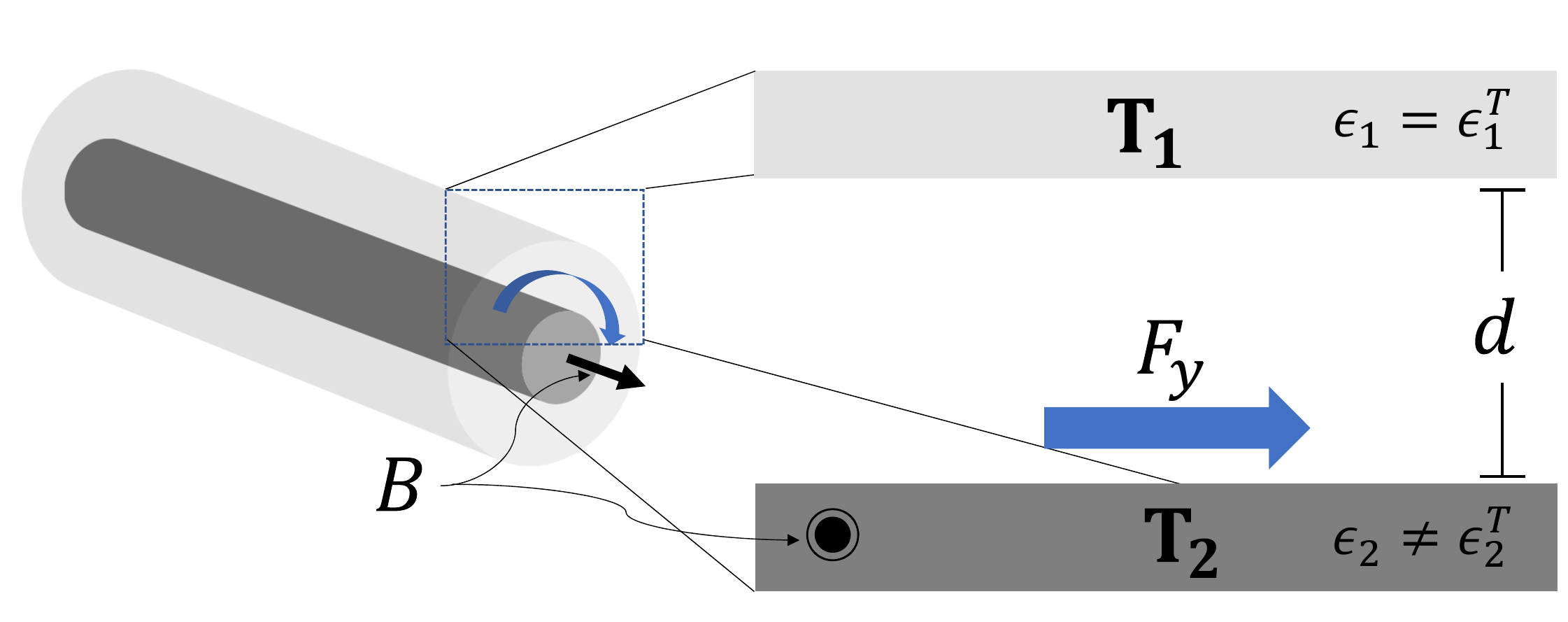}
    \caption{Two parallel plates at a distance $d$, at different temperatures.  At close proximity, these may mimic, e.g., inner and outer parts of an engine axis (left). We consider the case where the lower plate is non-reciprocal, which may, e.g., be due to a magnetic field pointing in the direction as indicated in the figure.}
    \label{fig:2}
\end{figure}

To find a bound in the near field, we consider two parallel, semi-infinite plates normal to $z$, separated by a distance $d$, as shown in Fig.~\ref{fig:2}. 
The force $F_y$ in Eq.~\eqref{eq:sforcetr3} can easily be extended to include nonzero $T_2$,
\begin{align}
F_y =\frac{2\hbar}{\pi} \int_0^\infty d\omega (n_1-n_2) \int \frac{dk_y}{2\pi} k_y S(k_y),\label{eq:sforcetr4}
\end{align}
with  $n_i=(e^{\frac{\hbar \omega}{k_B T_i}}-1)^{-1}$.  Here,
 $S(k_y)$ is the heat flux between  
 the two plates of surface area $A$, given by
~\cite{moncada2015magnetic},
\begin{align}
\frac{S(k_y)}{A}&\!=\!\int\!\frac{dk_x}{8\pi}\notag \biggl\{\mbox{Tr}_p\left[(1- \mathbbm{r}_2^{\dagger}\mathbbm{r}_2) \mathbb{D}(1-\mathbbm{r}_1\mathbbm{r}_1^{\dagger})\mathbb{D}^{\dagger}
\right]\Theta_{pr}\\
&+e^{-2|k_z| d}\mbox{Tr}_p\left[(\mathbbm{r}_2^{\dagger}\!-\!\mathbbm{r}_2) \mathbb{D}(\mathbbm{r}_1\!-\!\mathbbm{r}_1^{\dagger})\mathbb{D}^{\dagger}
\right]\Theta_{ev}\biggr\}, \label{eq:S1}
\end{align}
where $\mathbb{D}=(1-\mathbbm{r}_1 \mathbbm{r}_2 e^{2ik_z d})^{-1}$ and  $\mathbbm{r}_i$ is the  Fresnel reflection tensor of plate $i$: a $2 \times 2$ matrix in the space of polarizations~\cite{biehs2020near} with the trace taken in that space. We have introduced projectors $\Theta_{pr}$ and $\Theta_{ev}$
for propagating and evanescent modes, respectively.  Equation.~\eqref{eq:S1} holds for reciprocal or nonreciprocal plates, and in general $\mathbbm{r}_i$ has off-diagonal elements that couple the polarizations.

The constraints for motion in asymmetric ratchet systems out of equilibrium has been a subject of intense research in recent years \cite{denisov2014tunable,liao2020rectification,seifert2012stochastic,mogilner1996cell,julicher1997modeling,reimann2002brownian,reichhardt2017ratchet}. (The Feynman ratchet does not rotate in equilibrium, but does when heated \cite{feynman2011feynman}.) We may thus expect that if the plates in Fig.~\ref{fig:2} are not 
symmetric under $y\to -y$, e.g., for a tensorial dielectric response with tilted axes, a non-equilibrium heat flux will be generically accompanied by a ratchet-like \EPF. Surprisingly, this is not the case for asymmetric, but reciprocal plates:  As shown in Ref.~\cite{fan2020nonreciprocal} by explicit manipulations of Eq.~\eqref{eq:S1}, $S(k_y)$ is symmetric in $k_y$ if the two plates are made of reciprocal materials.
We extended this unexpected observation to arbitrary objects that are translationally invariant in direction $y$ \SI, and conclude that non-reciprocity is necessary for near field \EPF.
Furthermore, translational symmetry is an important element:
If object 1 is not translationally invariant, Eq.~\eqref{eq:sforcetr3} can yield a finite result for reciprocal media, as was found for an ellipsoid near a planar surface~\cite{Muller15}.

Assuming that at least one of the plates is nonreciprocal, we investigate the distance dependence of the force. For small separations, the integral over $k_x$ and $k_y$ in Eq.~\eqref{eq:sforcetr4} is dominated by the evanescent sector for large values of $k_\perp=\sqrt{k_x^2+k_y^2}\sim 1/d$.
We can thus expand the reflection tensors for large $k_\perp$.
In particular, we consider a dielectric tensor of the form
\begin{equation}
\bbeps=\left(\begin{array}{ccc}
\epsilon_{p} & 0 & 0\\
0 & \epsilon_{d} & -i\epsilon_{f}\\
0 & i\epsilon_{f} & \epsilon_{d}
\end{array}\right),\label{eq:permit}
\end{equation}
which is realized in magneto-optical materials with a dc magnetic field pointing along the $x$ direction~\cite{ishimaru2017electromagnetic}.
For $k_\perp^2\gg \frac{\omega^2}{c^2}|\varepsilon_{ij}|$, the reflection tensor becomes diagonal, and is dominated by the entry for electric polarization, $r^{NN}$. Denoting, in this limit, $r^{NN}\equiv r^\infty$, we find (using $k_y=k_\perp\sin\theta$)  \SI
\begin{align}
r^{\infty}=\frac{ (\epsilon_d-1)+ \sin\theta \epsilon_f }{ (\epsilon_d+1)+ \sin\theta \epsilon_f }.\label{eq:ri}
\end{align}
 By making $k_\perp$ dimensionless, $\tilde k_\perp= k_\perp d$, in the limit where $d$ is small compared to all other length scales, such as the thermal wavelength, and the material skin depth, we  obtain
\begin{align}
\frac{F_y}{A}&=\frac{2\hbar}{\pi d^3}\int_{0}^{\infty}d\omega (n_1-n_2)\notag\\ &\int\frac{d^{2}\tilde k_{\perp}}{(2\pi)^{2}}\tilde k_{y}
\frac{e^{-2|\tilde k_\perp|}\Im[r^\infty_2]\Im[r^\infty_1]}{|1-r_1^\infty r_2^\infty e^{-2|\tilde k_\perp|}|^2}.\label{eq:d3}
\end{align}
The force in Eq.~\eqref{eq:d3} is of similar form as normal forces of thermal origin \cite{antezza2008casimir}, and we expect it to be of experimental relevance, depending on materials used \SI; it diverges as $d^{-3}$ for small separations $d$. The well-known fact that heat transfer $H$ diverges as $d^{-2}$ in the same limit, which can easily be confirmed here by using the same rescaling with Eq.~\eqref{eq:ri}, implies that the ratio between force and heat transfer is proportional to $d^{-1}$. To quantify this, we compute $f(\omega,\theta)$ and $h(\omega,\theta)$, the force and heat transfer per frequency and per angle $\theta$ in the $k_xk_y$-plane. The integral over $|k_\perp|$ can then be performed to yield
\begin{align}
f(\omega,\theta)c = h(\omega,\theta)\frac{c}{\omega d}\sin\theta g[r_1^\infty(\theta) r_2^\infty(\theta)],\label{eq:b1}
\end{align}
where $g(x)$ is a well behaved positive function, which approaches unity for $|x|\to 0$, which corresponds to dilute materials \footnote{In the opposite limit of  $|x|\to \infty$, $g(x)\to\frac{\log|x|}{2}$}.

Equation~\eqref{eq:b1} may also be understood from Eq.~\eqref{eq:bound}: The integral of $\tilde k_y$ is limited by the exponential terms, implying $\tilde k_y^{max}\approx 1$. 

Using the mediant inequality yields, after integration over $\theta$, a bound valid for small distance $d$: in terms of the maximum of $|\sin\theta|g$ as a function of $\theta$,
\begin{align}
\frac{|f(\omega)|c}{ h(\omega)\mbox{max}(|\sin\theta| g)} \leq \frac{c}{\omega d}.\label{eq:lim}
\end{align}
Equation~\eqref{eq:lim} is a notable extension of Eq.~\eqref{eq:ff}. In the near field, the force is also bounded by the heat transfer, but the bound diverges as $d\to0$, since there is  no longer any energy--momentum relation constraining evanescent waves.
\begin{figure}
    \centering
    \includegraphics[width=0.5\textwidth]{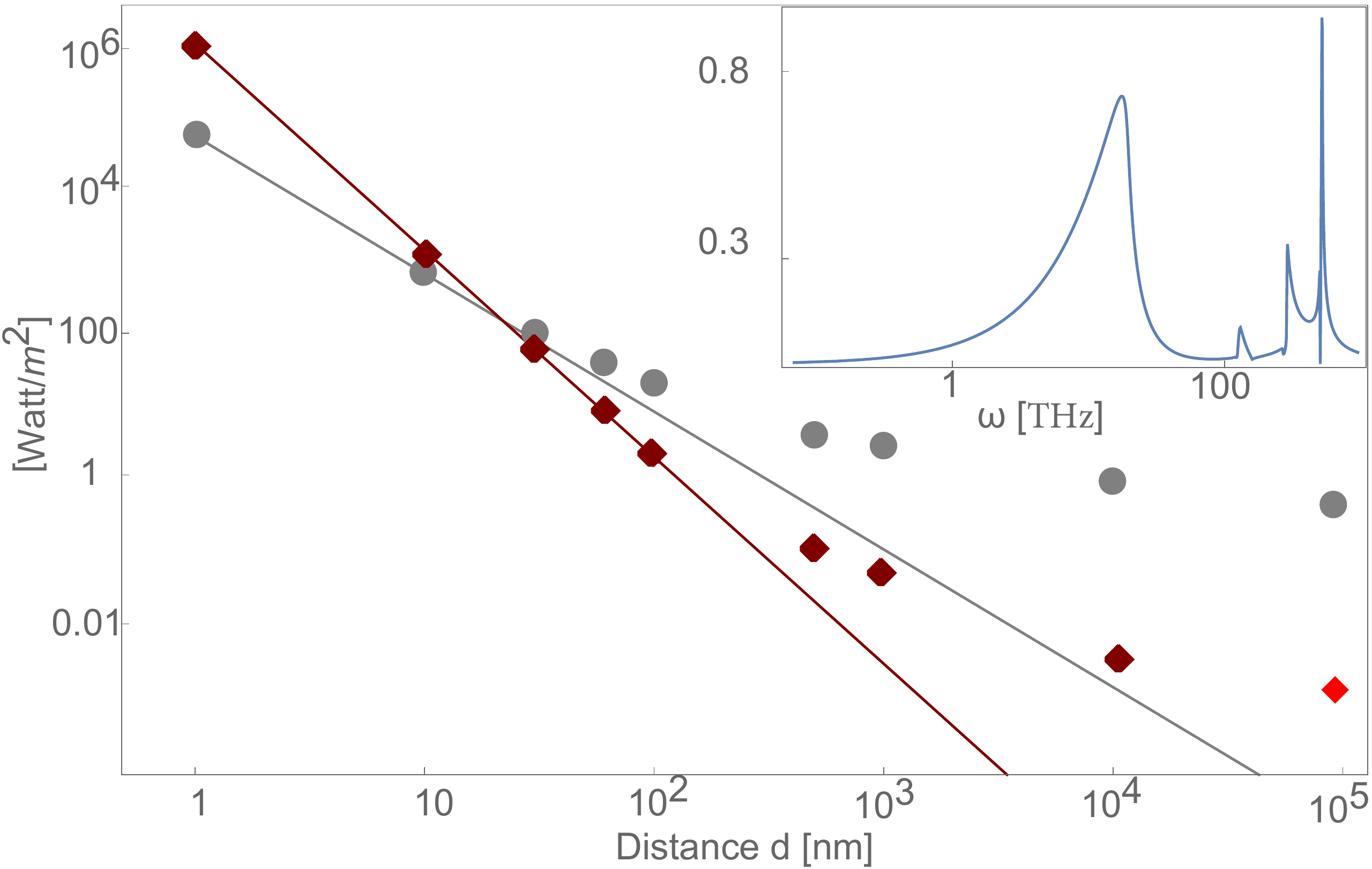}
    \caption{Scaled motive force $F_y c$ (red diamonds), and heat transfer $H$ (gray circles) for a SiC plate and one of n-InSb subject to a magnetic field along the $x$ axis. The dots correspond to numeric calculations and the continuous lines to the small $d$ asymptotes from Eq.~\eqref{eq:d3} and its equivalent for $H$. Note that the force changes sign from +y to -y (dark red to light red diamond) at large separation. Inset: l.h.s of Eq.~\eqref{eq:lim} in units of $c/(\omega d)$ for $d=1$ nm. Figures parameters: $B=10T$, $T_1=300K$, $T_2=270K$.}
    \label{fig:num}
\end{figure}

As a particular example, we consider that plate 1 is  composed of the reciprocal material SiC (with dielectric properties taken from Ref.~\cite{Spitzer59}), and   plate 2 made of n-doped  InSb under the influence of a magnetic field along the $x$ axis. The entries of $\bbeps$ in Eq.~\eqref{eq:permit} are given by~\cite{zhu2016persistent} $\epsilon_d=1-\frac{\omega_p^2(1+\frac{i\omega_\tau}{\omega})}{(\omega+i\omega_\tau)^2-\omega_b^2}$, $\epsilon_p=1-\frac{\omega_p^2}{\omega(\omega+i\omega_\tau)}$ and $\epsilon_f=-\frac{\omega_b \omega_p^2}{\omega((\omega+i\omega_\tau)^2-\omega_b^2)}$.

Here, $\omega_p$ is the plasma frequency and $\omega_\tau$ describes relaxation effects in InSb; the non-reciprocity ($\epsilon_f\neq0$) due to the magnetic field is encoded via the cyclotron frequency $\omega_b$.

Using material parameters tabulated in the SI, Fig.~\ref{fig:num} depicts $F_y c$ and $H$ plotted as functions of separation $d$.
As required by Eq.~\eqref{eq:bound}, the magnitude of $F_y c$ is smaller than $H$ in the far field. For distances below $\sim100$ nm, a steeper divergence of $F_y$ is observed, so that the two quantities cross at roughly 30 nm. For smaller separations, the on-shell energy--momentum relation of photons is noticeably broken.
This crossing point depends on the nonreciprocal properties of the materials, which in turn depend on the strength of the magnetic field\footnote{The force in Fig.~\ref{fig:num} is found to change sign at a distance of several tens of microns, an effect we leave to future work.}. The inset of Fig.~\ref{fig:num} shows the left-hand side of Eq.~\eqref{eq:lim} for the given materials as a function of frequency. It confirms  the inequality of Eq.~\eqref{eq:lim}, and that it presents a realistic bound, with some points reaching close to the bound. 

Acting as a heat engine in a setup such as in Fig.~\ref{fig:2},
the force $F_y$ can be used to extract work from the heat $H$ by 
moving the plate at a velocity $v$ along the $y$ direction.
The extracted power is given by $P=F_y v+\mathcal{O}(v^2)$, yielding 
to leading order in $v$, the efficiency
\begin{align}\label{eq:eta1}
\eta= \frac{F_y}{H} v +\mathcal{O}(v^2).
\end{align}

Prima facie, this expression suggests a Carnot efficiency, $\eta_c=\Delta T/T_1$  where $T_1-T_2=\Delta T>0$, at  $v_c=\frac{H}{F_y}\eta_c$, and exceeding 
it for larger velocities. However, as  shown below, for $v\sim v_c$ one cannot neglect $\mathcal{O}(v^2)$ terms.  Nonetheless, the first-order analysis provides an estimate  of the scale at which  efficiency is maximal.  In the far field,  Eq.~\eqref{eq:ff} implies that $v_c$ is larger than  $\eta_c c$, while in the near field, Eq.~\eqref{eq:lim} shows that it can be much smaller, going to zero linearly with $d$.

When running at velocities $v\sim v_c$, both force and heat exchange become functions of velocity, so that  higher order terms in Eq.~\eqref{eq:eta1} arise. 
While computing them poses no problem in principle, we restrict here to the limit of small $\Delta T\ll T_1$, where Onsager symmetry relations allow for some insights. While these relations rely on time-reversal symmetry in general, we find them to also be valid for nonreciprocal plates, leading to a modification of heat transfer between the plates, as
\begin{align}\label{eq:Onsager}
 {\cal {H}}\!=\!H\!+\!\frac{T_1}{\Delta T} F_y v+\mathcal{O}(v^2)\!=\!H\!
 \left[1+\frac{v}{v_c}+\mathcal{O}(v^2)\right]. 
\end{align}
The Onsager's relation leading to  Eq.~\eqref{eq:Onsager} ensures that $\eta$ remains below the Carnot limit. Interestingly, Eq.~\eqref{eq:Onsager} indicates that the given setup can also act as a refrigerator~\cite{vdbroeck06}: A backward  velocity of order $-v_c$ can remove heat from the cold plate. 

There is also an $\mathcal{O}(v)$ reduction to \EPF from friction
due to fluctuations of the electromagnetic  field, reducing the 
engine power to $P=v[F_y  -\Gamma v+\mathcal{O}(v^2)]$, where we have introduced the  
 friction coefficient $\Gamma$, relating force and velocity at $\Delta T=0$. It is given by a generalization of the result for reciprocal media~\cite{Volokitin07,Golyk13}, as
\begin{align}
\frac{\Gamma}{A}&=-\frac{2\hbar}{\pi}\int_{0}^{\infty}d\omega \frac{\partial n_1}{\partial \omega}\int\frac{dk_y}{2\pi} k_{y}^2 S(k_y)\geq 0,\label{eq:ga}
\end{align}
with $S(k_y)$ given in Eq.~\eqref{eq:S1}, so that $\Gamma\geq 0$. Expanding Eq.~\eqref{eq:ga} for small $d$ as in  Eq.~\eqref{eq:d3}, and using Eq.~\eqref{eq:ri}, leads to a $1/d^4$ divergence. Comparing $\Gamma v$ to $F_y$, one may thus expect the velocity scale $v_c$ to reappear. 

As a simple illustration, let us consider two dilute materials with 
$r^{\infty}_1$, $r^{\infty}_2\to 0$, and assume that the response of the non-reciprocal
medium is focused at a single frequency $\bar\omega$, such that \footnote{$\tilde\delta$ is a small window function of finite height and integral of unity.}, 
\begin{align}\label{eq:rApprox}
\Im[r_2^\infty](\omega)= r_n (1+\alpha\sin\theta+\dots)\bar\omega{\tilde\delta}(\omega-\bar\omega).
\end{align}
The parameter $\alpha$ is a dimensionless measure of influence of non-reciprocity on the reflection coefficient, and $r_n$ a unitless prefactor; 
the  requirement that the material be passive implies  $\Im[r^{\infty}_2]\geq0$,
and hence
$|\alpha|\leq1$ in absence of higher-order terms. 
This form yields
\begin{equation}\label{eq:FGH}
\frac{F_y}{A}=C\frac{\alpha\eta_c\hbar\bar\omega}{d^3}\,,\,\frac{H}{A}=2C\frac{\eta_c\hbar\bar\omega^2}{d^2}\,,\, \frac{\Gamma}{A}=\frac{3}{2}C\frac{\hbar}{ d^4}\,,
\end{equation}
where
$C=-\frac{\bar \omega}{8\pi^2} \frac{\partial n_1}{\partial \bar \omega}r_n\Im[r_1^\infty]$.
This leads to a velocity scale $v_c=\eta_c(2\bar\omega d/\alpha)$,
and an efficiency
\begin{equation}\label{eq:eff}
\eta=\eta_c{\frac{v}{v_c}}\frac{1-3{v}/({\alpha^2 v_c})+\cdots}{1+v/v_c+\cdots}\,.
\end{equation}
We expect higher-order terms in Eq.~\eqref{eq:Onsager} to become important
for $v\sim\bar\omega d$. However, the velocity scale $v_c$ carries an
additional factor of $\eta_c$, which allows us to ignore higher order
terms for $\Delta T\ll T_1$.
Equation~\eqref{eq:eff} then gives that velocity at maximum power as $v_{MP}=\alpha^2 v_c/6=\alpha\eta_c\bar\omega d/3$, at which point the efficiency is $\eta_{MP}=\eta_c{\frac{\alpha^2}{2(6+\alpha^2)}}$.

In summary, we have shown that for the case of two smooth plates at different temperatures,  non-equilibrium fluctuations  will cause a motive force if at least one of the plates is made of a nonreciprocal material. Although for large plate separation the on-shell energy-momentum photon relation limits the ratio between motive force and heat transfer, at small separations, the heat transfer and propulsion force are dominated by evanescent modes, allowing this ratio to grow inversely with the distance between the two plates. The velocity scale corresponding to maximal efficiency and power is thus linear in that distance.

\emph{Acknowledgments}
It is a pleasure to thank G.\ Bimonte, T.\ Emig, and R.\ L.\ Jaffe for helpful conversations.
D.\ G.\ K. is supported by the Gordon and Betty Moore Foundation as Physics of Living Systems Fellows through grant number GBMF4513.
N.\ G.\ was supported in part by the National Science Foundation (NSF) through grant PHY-1820700. M.K. acknowledges support from NSF through grant No. DMR-1708280.

%
\newpage
\onecolumngrid
\section*{Supplemental material}
\subsection{Derivation of Eq. (1)}
The non-equilibrium part of the force, acting on object 2 in the scenario depicted in Fig. 1 of the main text, is (generalization of the relations given in Ref.~\cite{kruger2012trace} to non-reciprocal cases), 
\begin{align}
\vct{F}&=\frac{2\hbar}{\pi} \int_0^\infty d\omega \frac{1}{e^{\frac{\hbar\omega}{k_BT_1}}-1}\Re \mbox{Tr} \left\{\boldsymbol{\nabla}(1+\mathbb{G}_0\mathbb{T}_{2})\frac{1}{1-\mathbb{G}_0\mathbb{T}_1\mathbb{G}_0\mathbb{T}_{2}}\mathbb{G}_0\left[\frac{\mathbb{T}_1-\mathbb{T}_1^\dagger}{2i} - \mathbb{T}_1 \Im[\mathbb{G}_0]\mathbb{T}_1^\dagger\right]\mathbb{G}^*_0 \frac{1}{1- \mathbb{T}^\dagger_2 \mathbb{G}^*_0\mathbb{T}^\dagger_1\mathbb{G}^*_0} \mathbb{T}_2^\dagger  \right\}.\label{eq:forceta}
\end{align}
We will continue by considering the force in $y$ direction and regard the case where $\mathbb{T}_2$ does not explicitly depend on $y$, so that 
\begin{align}
\partial_y \mathbb{T}_2=\mathbb{T}_2\partial_y.
\end{align}
We use that, if $B=B^\dagger$, ones has $\Re \mbox{Tr}[AB]=\frac{1}{2}\mbox{Tr}[(A+A^\dagger)B]$, and replace $\mathbb{T}_2^\dagger\partial_y(1+\mathbb{G}_0\mathbb{T}_{2})$ by its Hermitian part $\frac{1}{2}(\mathbb{T}_2^\dagger\partial_y-\partial_y\mathbb{T}_2)+i\partial_y\mathbb{T}_2^\dagger\Im[\mathbb{G}_0]\mathbb{T}_2$,
{\footnotesize
\begin{align}
F_{y}
&=\frac{-2\hbar}{\pi} \int_0^\infty \!\!\!\!\! d\omega \frac{1}{e^{\frac{\hbar\omega}{k_BT_1}}-1} \mbox{Tr} \left\{i\partial_y\left[\frac{\mathbb{T}_2-\mathbb{T}_2^\dagger}{2i}-\mathbb{T}_2^\dagger\Im[\mathbb{G}_0]\mathbb{T}_{2}\right]\frac{1}{1-\mathbb{G}_0\mathbb{T}_1\mathbb{G}_0\mathbb{T}_{2}}\mathbb{G}_0\left[\frac{\mathbb{T}_1-\mathbb{T}_1^\dagger}{2i} - \mathbb{T}_1 \Im[\mathbb{G}_0]\mathbb{T}_1^\dagger\right]\mathbb{G}^*_0 \frac{1}{1- \mathbb{T}^\dagger_2 \mathbb{G}^*_0\mathbb{T}^\dagger_1\mathbb{G}^*_0} \right\}
\end{align}}
This is Eq. (1) in the main text.

\subsection{Positivity of $\hat{\mathbb{R}}_2$}
We consider an object with $\mathbb{V}=\tilde{\mathbb{V}}(x,x',z,z') \delta(y-y')$. We can write, using an expansion in Fourier modes along $y$,
\begin{align}
\frac{\mathbb{V}-\mathbb{V}^\dagger}{2i}=\int \frac{dk_y}{2\pi} \frac{\tilde{\mathbb{V}}-\tilde{\mathbb{V}}^\dagger}{2i}e^{ik_y (y-y')}\label{eq:la}
\end{align}
Importantly, we have $l\mbox{Tr}[\mathbb{V}-\mathbb{V}^\dagger]=\mbox{Tr}[\tilde{\mathbb{V}}-\tilde{\mathbb{V}}^\dagger]$, where $l$ is a length scale that formally goes to zero. It can be understood when introducing an upper cut off $\sim{1/l}$ in Eq.~\eqref{eq:la}. We thus have  $\mbox{Tr}[(\tilde{\mathbb{V}}-\tilde{\mathbb{V}}^\dagger)/2i]\geq 0$.

Furthermore, for such potential, $\mathbb{G}=\mathbb{G}(y-y')$, so that 
\begin{align}
\mathbb{G}&=\int \frac{dk_y}{2\pi} \tilde{\mathbb{G}}(k_y)e^{ik_y (y-y')}\\
\mathbb{G}^\dagger&=\int \frac{dk_y}{2\pi} [\tilde{\mathbb{G}}(k_y)]^\dagger e^{ik_y (y-y')}.
\label{eq:G}
\end{align}
We finally have $\tilde{\mathbb{R}}_2$ defined in the main text,

\begin{align}
    \tilde{\mathbb{R}}_2=\tilde{\mathbb{G}} \frac{\tilde{\mathbb{V}}-\tilde{\mathbb{V}}^\dagger}{2i} \tilde{\mathbb{G}}^\dagger\geq 0.
\end{align}
And 
\begin{align}
    \hat{\mathbb{R}}_2=e^{ik_yy}\tilde{\mathbb{R}}_2e^{-ik_y y'}\geq 0.
\end{align}

\subsection{Onsager relation and friction}
Generalizing the results of Ref.~\cite{Golyk13} for non-reciprocal objects yields the following result for the friction tensor, relating the force acting on object 2 in response to a velocity of object 1 (both for $y$-direction, both objects and environment have same temperature $T_2$) 
{\footnotesize
\begin{equation}\label{FF}
{\Gamma}_1^{(2)}=-\frac{2\hbar}{\pi}\int_{0}^{\infty}d\omega \frac{\partial n_2}{\partial \omega} \mbox{Tr} \left\{\partial^2_y\left[\frac{\mathbb{T}_2-\mathbb{T}_2^\dagger}{2i}-\mathbb{T}_2^\dagger\Im[\mathbb{G}_0]\mathbb{T}_{2}\right]\frac{1}{1-\mathbb{G}_0\mathbb{T}_1\mathbb{G}_0\mathbb{T}_{2}}\mathbb{G}_0\left[\frac{\mathbb{T}_1-\mathbb{T}_1^\dagger}{2i} - \mathbb{T}_1 \Im[\mathbb{G}_0]\mathbb{T}_1^\dagger\right]\mathbb{G}^*_0 \frac{1}{1- \mathbb{T}^\dagger_2 \mathbb{G}^*_0\mathbb{T}^\dagger_1\mathbb{G}^*_0} \right\}.\\
\end{equation}}
We used that both object are translationally invariant in direction $y$, so that $\partial_y$ can be moved through. We continue by regarding only effects for the inside between plates, for which forces are equal and opposite. The part given here ${\Gamma}_1^{(2)}=-\Gamma$ is thus the negative of $\Gamma$ in  Eq. (18). The relation between $F_y$ and $\Gamma$ given in the main text is apparent.

Similarly, the heat transfer absorbed by object two changes with $v$ in the following way,
{\footnotesize
\begin{align}
\left.\frac{\partial{\cal H}^{(2)}}{\partial v}\right|_{v=0}=-\frac{2\hbar}{\pi}\int_{0}^{\infty}d\omega \omega \frac{\partial n_2}{\partial \omega} \mbox{Tr} \left\{i\partial_y\left[\frac{\mathbb{T}_2-\mathbb{T}_2^\dagger}{2i}-\mathbb{T}_2^\dagger\Im[\mathbb{G}_0]\mathbb{T}_{2}\right]\frac{1}{1-\mathbb{G}_0\mathbb{T}_1\mathbb{G}_0\mathbb{T}_{2}}\mathbb{G}_0\left[\frac{\mathbb{T}_1-\mathbb{T}_1^\dagger}{2i} - \mathbb{T}_1 \Im[\mathbb{G}_0]\mathbb{T}_1^\dagger\right]\mathbb{G}^*_0 \frac{1}{1- \mathbb{T}^\dagger_2 \mathbb{G}^*_0\mathbb{T}^\dagger_1\mathbb{G}^*_0} \right\}.
\end{align}}
The force $F_y$ can be expanded in temperature difference (again using simplifications of symmetry arising in the  setup of two parallel plates). Observing $n_1-n_2=-\frac{\partial n_2}{\partial T_2}(T_2-T_1)+\dots=\frac{\omega}{T_2}\frac{\partial n_2}{\partial \omega}(T_2-T_1)+\dots$ yields Eq. (17) in the main text.

\subsection{Derivation of Eq. (11)}
We calculate $r^{NN}$ at $k_{\perp}\gg \frac{\omega}{c}|\epsilon_{ij}|$ for  an interface between the vacuum and a plane with normal vector parallel to $\hat{z}$. $r^{NN}=\frac{n_r}{n_i}$ where $n_r$ is the amplitude of the  electric field of the reflected wave, $\mathbf{E_r}=n_r\frac{c}{\omega}\{k_z k_x,k_zk_y,k^2_{\perp}\}$ and $n_i$ is the amplitude of the electric field of the incoming wave, $\mathbf{E_i}=n_i\frac{c}{\omega}\{-k_z k_x,-k_zk_y,k^2_{\perp}\}$. $\mathbf{k_i}$ and $\mathbf{k_r}$ are the corresponding wave vectors. At this point we assume that there could be two different transmitted waves \cite{chen1981coordinate} with wave vectors $\mathbf{k_+}$ and $\mathbf{k_-}$ and electric fields $\mathbf{E_+}$ and $\mathbf{E_-}$ respectively.  For $k_{\perp}\gg \frac{\omega}{c}|\epsilon_{ij}|$ the $z$-component of the transmitted wave vectors can be approximated as  $i k_{\perp}$ and  $ \sqrt{-k_y^2-k_x^2\frac{\epsilon_p}{\epsilon_d}} $. Using the fact that  the wave vector is perpendicular to the electric field, $\mathbf{k_+}\cdot\mathbf{E_+}=\mathbf{k_-}\cdot \mathbf{E_-}=0$ with  the standard boundaries conditions  that are used  for calculating the Fresnel coefficients, that is:
\begin{gather}
    \left(\mathbf{E_i}+\mathbf{E_r}-\bbeps  \mathbf{E_+}-\bbeps\mathbf{E_-}\right)\cdot  \hat{z}=0 \\
     \left(\mathbf{k_i}\times\mathbf{E_i}+\mathbf{k_r}\times\mathbf{E_r}- \mathbf{k_+}\times\mathbf{E_+}-\mathbf{k_-}\times\mathbf{E_-}\right)\cdot  \hat{z}=0\\
     \left(\mathbf{E_i}+\mathbf{E_r}-  \mathbf{E_+}-\mathbf{E_-}\right)\times \hat{z}=0 \\
         \left(\mathbf{k_i}\times\mathbf{E_i}+\mathbf{k_r}\times\mathbf{E_r}- \mathbf{k_+}\times\mathbf{E_+}-\mathbf{k_-}\times\mathbf{E_-}\right)\times  \hat{z}=0
    \end{gather}

one can find all the components of $ \mathbf{E_+}$ and $ \mathbf{E_-}$ and also Eq. (11) in the main text.

\subsection{Permittivities}

The permittivity of SiC is given by a diagonal matrix with diagonal entries 
\begin{equation}
    \epsilon_{SiC}=\epsilon_{\infty}\frac{\omega^2-\omega^2_{LO}+i\omega \gamma}{\omega^2-\omega^2_{TO}+i\omega \gamma}
\end{equation}
where $\epsilon_{\infty}=6.7$, $\omega_{LO}=0.12eV$, $\omega_{TO}=0.098 eV$ and $\gamma=5.88\times10^{-4}eV$.

The permittivity matrix for the n-doped InSb  is given in the main text. The plasma frequency is 
$\omega_p=0.51eV$, the cyclotron frequency is $\omega_{b}=0.0153eV$ (for  $B=10T$) and  $\omega_{\tau}=4.1357\times10{-3}eV$.

\subsection{Comparison with experimental capabilities}
We estimate that the  propulsion force magnitude is within the accuracy of current experimental capabilities, that is $\sim 1$mPa for an object separation of $d\sim 100$nm \cite{decca2007tests}. This can be achieved by considering that both plates have  the same reflection coefficients and changing the material doping in order to
change the material plasma frequency\cite{law2014doped}. For $\omega_p=100THz$ and $d=100$nm the Casimir pressure is of the order of $10$mPa. The force can furthermore strongly be increased by using higher temperatures.

\subsection{Symmetry relations for heat transfer and consequences for motive forces}
\subsubsection{General}
We introduce the following exact expressions for the heat $H_i^{(2)}$ emitted by object $i$ (including environment) and absorbed by object 2. These expressions are generalizations of the ones given in Ref.~\cite{Kruger11}, and are valid for objects of arbitrary shapes and reciprocal or nonreciprocal media.  (with $\mathbb{R}_1$, $\mathbb{R}_2$ and $\mathbb{W}$ given in the main text)
\begin{align}
H_1^{(2)}(T)&= \frac{2\hbar}{\pi} \int_0^\infty d\omega \frac{\omega}{e^{\frac{\hbar\omega}{k_BT}}-1} \mbox{Tr} \left\{\mathbb{R}_2\mathbb{W}\mathbb{R}_1\mathbb{W}^\dagger\right\}\\
H_2^{(2)}(T)&=\frac{2\hbar}{\pi} \int_0^\infty d\omega \frac{\omega}{e^{\frac{\hbar\omega}{k_BT}}-1} \mbox{Tr} \left\{\frac{\mathbb{G}_1-\mathbb{G}^\dagger_1}{2i}\frac{1}{1-\mathbb{T}_2\mathbb{G}_0\mathbb{T}_1\mathbb{G}_0}\left[\frac{\mathbb{T}_2-\mathbb{T}_2^\dagger}{2i}-\mathbb{T}_2\Im[\mathbb{G}_0]\mathbb{T}_2^\dagger\right] \frac{1}{1-\mathbb{G}^*_0\mathbb{T}^\dagger_1\mathbb{G}^*_0\mathbb{T}_2^\dagger} \right\}\label{eq:em}\\&= \frac{-2\hbar}{\pi} \int_0^\infty d\omega \frac{\omega}{e^{\frac{\hbar\omega}{k_BT}}-1} \mbox{Tr} \left\{\mathbb{R}_2\mathbb{W}\frac{\mathbb{G}_1-\mathbb{G}^\dagger_1}{2i}\mathbb{W}^\dagger\right\}\label{eq:abs}\\
H_{env}^{(2)}(T)&= \frac{2\hbar}{\pi} \int_0^\infty d\omega \frac{\omega}{e^{\frac{\hbar\omega}{k_BT}}-1} \mbox{Tr} \left\{\mathbb{R}_2\mathbb{W}(1+\mathbb{G}_0\mathbb{T}_1)\Im[\mathbb{G}_0] (1+\mathbb{T}_1^\dagger\mathbb{G}_0^*)\mathbb{W}^\dagger\right\}
\end{align}
The equality of Eqs.~\eqref{eq:em} and \eqref{eq:abs} can be shown in a tedious, but straightforward calculation. (Note that the operator in square brackets of Eq.~\eqref{eq:em} is different from $\mathbb{R}_2$.)
We now transform all operators in Fourier space along $y$,
\begin{align}
     \tilde{\mathbb{O}}(k_y)&=\int d{y} e^{-ik_y y}  \mathbb{O}(y)\\
     \tilde{\mathbb{O}}^\dagger(k_y)&=\int d{y} e^{-ik_y y}  \mathbb{O}^\dagger(y)
\end{align}
So that
\begin{align}
\mbox{Tr} \left\{\mathbb{R}_2\mathbb{W}\mathbb{R}_1\mathbb{W}^\dagger\right\}&= \int \frac{dk_y}{2\pi}\mbox{Tr} \left\{e^{-ik_y y} {\tilde{\mathbb{R}}}_2\tilde{\mathbb{W}}\tilde{\mathbb{R}}_1\tilde{\mathbb{W}}^\dagger e^{ik_y y'} \right\}= \int \frac{dk_y}{2\pi} S_1^{(2)}(k_y)\\
\mbox{Tr} \left\{\mathbb{R}_2\mathbb{W}\frac{\mathbb{G}_1-\mathbb{G}^\dagger_1}{2i}\mathbb{W}^\dagger\right\}&= \int \frac{dk_y}{2\pi}\mbox{Tr} \left\{e^{-ik_y y} {\tilde{\mathbb{R}}}_2\tilde{\mathbb{W}}   \frac{\tilde{\mathbb{G}}_1-\tilde{\mathbb{G}^\dagger_1}}{2i}\tilde{\mathbb{W}}^\dagger e^{ik_y y'} \right\}= -\int \frac{dk_y}{2\pi} S_2^{(2)}(k_y)\\
\mbox{Tr} \left\{\mathbb{R}_2\mathbb{W}\mathbb{G}_1\Im[\mathbb{G}_0] \mathbb{G}^\dagger_1\mathbb{W}^\dagger\right\}&= \int \frac{dk_y}{2\pi}\mbox{Tr} \left\{e^{-ik_y y} {\tilde{\mathbb{R}}}_2\tilde{\mathbb{W}}   \tilde{\mathbb{G}}_1\frac{\tilde{\mathbb{G}}_0-\tilde{\mathbb{G}^\dagger_0}}{2i}\tilde{\mathbb{G}}^\dagger_1\tilde{\mathbb{W}}^\dagger e^{ik_y y'} \right\}= \int \frac{dk_y}{2\pi} S_{env}^{(2)}(k_y)
\end{align}
we note that all operators, even if reciprocal in real speace, may not be reciprocal for mode $k_y$, such as $\tilde{\mathbb{G}}_0$.

\subsubsection{Absence of equilibrium transfer per wavevector for reciprocal and nonreciprocal media}
Another tedious, but straight forward computation shows that
\begin{align}
\sum_{i=1,2,env}S_i^{(2)}(k_y)=0.
\end{align}
This shows that, in equilibrium, no object heats up or cools down, {\it per wavevector $k_y$}. One may interpret that the second law of thermodynamics holds per wavevector $k_y$ \cite{fan2020nonreciprocal}. This statement, which holds for reciprocal and non-reciprocal objects, was proven for two parallel plates in \cite{fan2020nonreciprocal}. Notably, the generalization for arbitrary objects involves the environment, and does not (at least not obviously) hold pairwise for the two objects involved. In case that radiation to or from the environment can be neglected, we have
\begin{align}
S_1^{(2)}(k_y)=-S_2^{(2)}(k_y)=S_2^{(1)}(k_y).\label{eq:2nd}
\end{align}
In the last equality, we used conservation of energy for two objects.

\subsubsection{Detailed balance per wavevector for reciprocal media}
For reciprocal media, all operators involved are symmetric. For such symmetric, translationally invariant operator $\mathbb{O}$ (free Green's function or $\mathbb{T}$), one has
\begin{align}
     \tilde{\mathbb{O}}(k_y)&=\int d{y} e^{-ik_y y}  \mathbb{O}(y)
     =\int d{y} e^{ik_y y}  \mathbb{O}(-y)
     =\int d{y} e^{ik_y y}  \mathbb{O}^T(y)
     =\tilde{\mathbb{O}}^T(-k_y).
\end{align}
We introduce slightly different notation compared to the main text: The emission operator  
$\mathbb{M}_i=\mathbb{G}_0\left[\frac{\mathbb{T}_i-\mathbb{T}_i^\dagger}{2i} - \mathbb{T}_i \Im[\mathbb{G}_0]\mathbb{T}_i^\dagger\right]\mathbb{G}^*_0$ and the absorption operator $\mathbb{ {A}}_i=\mathbb{G}_0^*\left[\frac{\mathbb{T}_i-\mathbb{T}_i^\dagger}{2i} - \mathbb{T}_i^\dagger \Im[\mathbb{G}_0]\mathbb{T}_i\right]\mathbb{G}_0$
and  the multiple scattering operator $
\mathbb{W}_{ij}= \mathbb{G}_0^{-1}(1-\mathbb{G}_0\mathbb{T}_i\mathbb{G}_0\mathbb{T}_{j})^{-1}$.
We have, in Fourier space
\begin{align}
\tilde{\mathbb{M}}_i^*(-k_y)&=\tilde{\mathbb{A}}_i(k_y)\\
\tilde{\mathbb{A}}_i^*(-k_y)&=\tilde{\mathbb{M}}_i(k_y)\\
\tilde{\mathbb{W}}_{21}^*(-k_y)&=\tilde{\mathbb{W}}_{12}^\dagger(k_y)\\
(\tilde{\mathbb{W}}_{21}^\dagger)^*(-k_y)&=\tilde{\mathbb{W}}_{12}(k_y)
\end{align}
Using it, and the fact that the expression is real, we have
\begin{align}
    S_1^{(2)}(k_y)=\mbox{Tr}\left\{\tilde{\mathbb{A}}_2(k_y)\tilde{\mathbb{W}}_{12}(k_y)\tilde{\mathbb{M}}_1(k_y)\tilde{\mathbb{W}}^\dagger_{12}(k_y)\right\}=\mbox{Tr}\left\{\tilde{\mathbb{M}}_2(-k_y)\tilde{\mathbb{W}}^\dagger_{21}(-k_y)\tilde{\mathbb{A}}_1(-k_y)\tilde{\mathbb{W}}_{21}(-k_y)\right\}
    =S_2^{(1)}(-k_y).\label{eq:db}
\end{align}
\subsubsection{Consequences for forces}
If radiation from environment can be neglected, for reciprocal media, we have, combining Eqs.~\eqref{eq:2nd} and \eqref{eq:db},
\begin{align}
S_1^{(2)}(k_y)=S_1^{(2)}(-k_y).
    \end{align}
    Using Eq.~(3) in the main text shows that in this case, the motive force vanishes. Motive forces for reciprocal media, in a setting of translational invariance along $y$, can only, if at all, result from far field radiation from the environment. It is thus not possible to obtain near field propulsion forces for nonreciprocal media.   

\end{document}